High Thermoelectric Figure of Merit by Resonant Dopant in Half-Heusler Alloys

Long Chen,[1,a)] Yamei Liu,[2] Jian He,[2] Terry M. Tritt,[2,3] and S. Joseph Poon[1,a)]

[1] *Department of Physics, University of Virginia, Charlottesville, Virginia 22904-4714*

[2] *Department of Physics and Astronomy, Clemson University, Clemson, South Carolina  29634-0978*

[3] *Materials Science & Engineering Department, Clemson University, Clemson, South Carolina  29634*

Abstract:

Half-Heusler alloys have been one of the benchmark high temperature thermoelectric materials owing to their thermal stability and promising figure of merit *ZT*. Simonson *et al.* early showed that small amounts of vanadium doped in $Hf_{0.75}Zr_{0.25}NiSn$ enhanced the Seebeck coefficient and correlated the change with the increased density of states near the Fermi level. We herein report a systematic study on the role of vanadium (V), niobium (Nb), and tantalum (Ta) as prospective resonant dopants in enhancing the *ZT* of *n*-type half-Heusler alloys based on $Hf_{0.6}Zr_{0.4}NiSn_{0.995}Sb_{0.005}$. The V doping was found to increase the Seebeck coefficient in the temperature range 300-1000 K, consistent with a resonant doping scheme. In contrast, Nb and Ta act as normal *n*-type dopants, as evident by the systematic decrease in electrical resistivity and Seebeck coefficient. The combination of enhanced Seebeck coefficient due to the presence of V resonant states and the reduced thermal conductivity has led to a state-of-the-art *ZT* of 1.3 near 850 K in *n*-type $(Hf_{0.6}Zr_{0.4})_{0.99}V_{0.01}NiSn_{0.995}Sb_{0.005}$ alloys.

[a)] Author to whom correspondence should be addressed; electronic mail: lc4wn@virginia.edu; sjp9x@virginia.edu

Thermoelectric materials can directly convert heat to electricity and vice versa, making thermoelectrics an important component of the renewable energy technology. The performance of thermoelectric material is measured by the dimensionless figure of merit $ZT$, $ZT = (S^2\sigma/\kappa)T$, where $\sigma$ is the electrical conductivity, S is the Seebeck coefficient, $\kappa$ is the total thermal conductivity. Among the state-of-the-art thermoelectric materials, the RNiSn (R = Hf, Zr and Ti) half-Heusler (HH) phases have been noted by their thermal stability,[1,2] scalability, and potential for large power output,[3-6] and the combination of high Seebeck coefficient and low thermal conductivity.[7-12] Until recently, the highest verifiable $ZT$ value of HH alloys has been around 1 in the intermediate to high temperature range,[4,9,10] developing higher $ZT$ HH alloys in a challenging task.[2,6,13-16] Per the definition of $ZT$, there are two basic approaches to enhancing the $ZT$: (1) reducing the thermal conductivity, and (2) increasing the electrical conductivity and Seebeck coefficient. The present work is in line with the second approach and the focus is on enhancing the Seebeck coefficient via resonant doping.

After Mahan and Sofo's prediction, the highest ZT for a given material could be achieved when the electronic density of states resembled a Dirac delta function at the Fermi level.[17] In this vein, there are several works on implementing resonant states near the Fermi level.[18-22] Heremans *et al.*[18] observed a doubled ZT at near 800 K by doping 2% Tl to PbTe, and it was attributed to the distorted density of states at the valence band edge. Predictions from first-principles density functional theory calculations were performed in N-doped GaAs[19] and O-doped ZnSe systems.[20] A localized sub-band in the density of states was created by the dopants in both systems. Bilc *et al.* performed *ab initio* electronic structure calculations in $AgPb_mSbTe_{2+m}$ compositions, and showed the presence of resonant states.[21] For half-Heusler alloys, Simonson *et al.*[22] showed that light V doping on the group IVB metal site of $Hf_{0.75}Zr_{0.25}NiSn$ enhanced the Seebeck coefficient at temperatures up to 650 K, the enhanced Seebeck coefficient was correlated to the increased density of states near the Fermi level. In view of the recent success in synthesizing the half-Heusler alloys with $ZT > 1$, it is important from the thermoelectric materials research perspective to determine whether all VA group elements (V, Nb, and Ta) can enhance the $ZT$ of half-Heusler alloys. We report herein a systematic investigation of the role of V, Nb, and Ta as potential resonant dopants in the polycrystalline n-type half-Heusler alloys based on $Hf_{0.6}Zr_{0.4}NiSn_{0.995}Sb_{0.005}$. A high $ZT$ of 1.3 is obtained in n-type $(Hf_{0.6}Zr_{0.4})_{0.99}V_{0.01}NiSn_{0.995}Sb_{0.005}$.

Ingots of $(Hf_{0.6}Zr_{0.4})_{1-x}M_xNiSn_{0.995}Sb_{0.005}$ (M= V, Nb and Ta) with x= 0%, 0.2%, 0.5% and 1.0% were arc melted from appropriate quantities of elemental Hf, Zr, M, Sn and pre-melted $Sn_{90}Sb_{10}$ alloy under argon atmosphere. The Sn-Sb precursor alloy instead of Sb element was used in view of the small amount of Sb in the alloys and low sublimation point of Sb. Then ingots were pulverized into 10 to 50 μm size pieces followed by consolidation using Spark Plasma Sintering (Thermal Technologies® SPS 10-4) technique. The samples were first sintered at a lower temperature of 1300 °C for 10 mins under 60 MPa to ensure single phase in the mixed-phase ingots. Then, these samples were annealed at a higher temperature of 1350 °C for another 20 mins. The crystal structures of the samples were characterized by x-ray diffraction on the PANalytical® X'Pert Pro MPD (Multi Purpose Diffractometer) instrument. Energy-dispersive spectroscopy (EDS) was performed to check the distribution of the constituent elements. The resistivity and thermopower were measured using ULVAC® ZEM 3 system. The thermal conductivity was calculated from the sample density D, the specific heat $C_p$ (from Netzsch® Differential Scanning Calorimeter), and the thermal diffusivity α (from Netzsch® LFA 457 MicroFlash system) as $\kappa = DC_p\alpha$. The electrical thermal conductivity $\kappa_e$ was estimated by using the Wiedenann-Franz relationship $\kappa_e = L\sigma T$, where L is the Lorenz Number which can be determined by using the equation proposed by Kim et al.[23] The lattice thermal conductivity $\kappa_L$ was achieved by subtracting the electrical thermal conductivity $\kappa_e$ from the total thermal conductivity κ.

The X-ray pattern of the sample that contained the highest vanadium content, $(Hf_{0.6}Zr_{0.4})_{0.99}V_{0.01}NiSn_{0.995}Sb_{0.005}$, is displayed in Figure 1. The pattern can be indexed to HH structure (space group F$\underline{4}$3m, Cl$_b$) and no peaks of a second phase are observed. The compositional homogeneity checked by EDS line scan shows a uniform distribution of vanadium throughout the microstructure, which is similar to that reported earlier.[22]

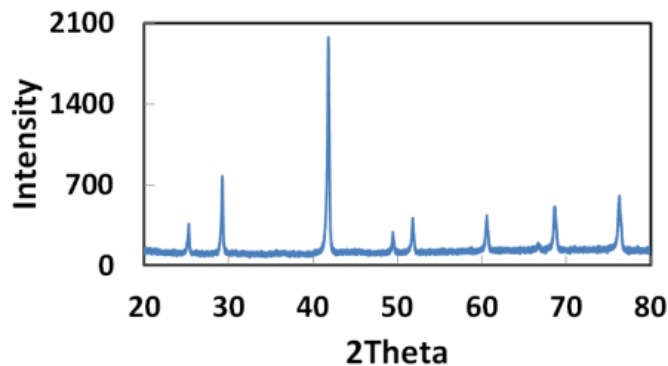

FIG. 1. X-ray pattern for n-type $(Hf_{0.6}Zr_{0.4})_{0.99}V_{0.01}NiSn_{0.995}Sb_{0.005}$.

The temperature dependencies of the electrical resistivity $\rho$, the Seebeck coefficient $S$, and the power factor are shown in Figure 2. It can be seen that as the doping level of V increases, the Seebeck coefficient increases over the entire temperature range 300 – 1000 K accompanied by an increase in the resistivity. In the maximal case of 1% V-doped sample, the Seebeck coefficient increases by around 10% compared with the undoped sample. Despite the increase in $S$, $PF = S^2/\rho$ of the sample with 1% V remains similar compared with that of the undoped sample due to the increase in $\rho$. In contrast, doping effects are different for Nb or Ta doped samples. Both the electrical resistivity and the Seebeck coefficient decrease systematically as more Nb or Ta is added. This evidence indicates that Nb and Ta work as normal *n*-type dopants. The noted different doping effects between 3d V dopant and same group 4d Nb and 5d Ta dopants can be qualitatively understood in terms of hybridization of Group 5 dopant atoms and Hf(Zr) host atoms. Hybridization of the valence electrons between the Hf(Zr) and X (X=V, Nb, and Ta) atoms via their nearest-neighbor interaction $\upsilon$ can cause bonding-antibonding splitting. Following the discussion in reference 22, the antibonding level lies above the Hf(Zr) atomic level $E_{Hf}$ by an amount $\upsilon^2/|E_X-E_{Hf}|$, where $E_X$ denotes the atomic level of X.[24] Since the atomic levels of Nb and Ta lie closer to the Hf level than that of V, the antibonding level that results from Hf-(Nb or Ta) hybridization is expected to lie somewhere further away from the conduction band edge. Moreover, based on the interatomic distance between Hf and X, the coupling matrix element $\upsilon$ would increase from V to Nb and Ta.[25] The weaker hybridization between V and Hf would also infer a more 'localized' nature of the V induced hybridized states near the conduction band edge, giving rise to the resonant states that increases the electrical resistivity and enhances the thermopower.

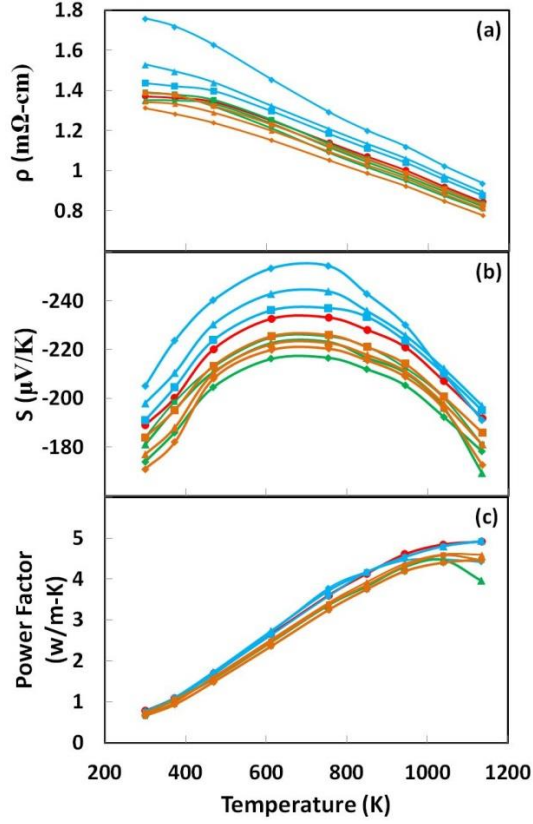

FIG. 2. Thermoelectric properties of n-type $(Hf_{0.6}Zr_{0.4})_{1-x}M_xNiSn_{0.995}Sb_{0.005}$ (M = V, Nb and Ta), where x = 0 (red circle), x = 0.002 with M = V (blue square), x = 0.005 with M = V (blue triangle), x = 0.01 with M = V (blue rhombus), x = 0.002 with M = Nb (green square), x = 0.005 with M = Nb (green triangle), x = 0.01 with M = Nb (green rhombus), x = 0.002 with M = Ta (orange square), x = 0.005 with M = Ta (orange triangle), x = 0.001 with M = Ta (orange rhombus): (a) Electrical resistivity ($\rho$), (b) Seebeck coefficient (S), and (c) power factor.

The Hall coefficient $R_H = -1/nq$ reveals both the carrier type and carrier concentration, where *n* is the carrier concentration and *q* is the carrier charge. The carrier mobility $\mu_H$ can be deduced from the relation ($\sigma = nq\mu_H$), where $\sigma$ is the electrical conductivity. The effective band mass is calculated by assuming a single-band model[26]:

$$S = \pm \frac{k_B}{e}\left[\frac{2F_1(\eta_F)}{F_0(\eta_F)} - \eta_F\right] \quad (1)$$

$$F_n(\eta_F) = \int_0^\infty \frac{x^n}{1+e^{(x-\eta_F)}}dx \quad (2)$$

$$n = \frac{4}{\pi}\left(\frac{2\pi m^* k_B T}{h^2}\right)^{\frac{3}{2}} F_{1/2}(\eta_F) \quad (3)$$

Where $F_n(\eta_F)$ is the Fermi-Dirac integral, $\eta_F$ is the reduced Fermi level defined as $\eta_F = E_F/k_B T$, $k_B$ is the Boltzmann constant, $m^*$ is the effective band mass. $h$ is the Planck constant, and $T$ is the temperature in K. The use of the single-band model for the above analysis was justified by previous study on the same system of materials.[22] The $n$, $\mu_H$, and calculated $m^*$ values are shown in Table I. It can be seen that n decreases for V doping and increases for Nb and Ta doping. The result further confirms the above-mentioned roles of the three dopants in the host alloy. The effective band mass for these alloys has been estimated to be larger than $2m_e$.[22,27] The calculated $m^*$ value increases from $2.16 m_e$ for undoped sample to $2.72 m_e$ for 1% V-doped sample, and slightly decreases for Nb and Ta-doped samples. This enhanced effective band mass might be attributed to a distortion in the density of states near the Fermi energy for V-doped samples.[22]

TABLE I. The room-temperature Seebeck coefficient, Hall coefficient ($R_H$), carrier concentration ($n$), Hall mobility ($\mu_H$), and effective band mass ($m^*$) of $(Hf_{0.6}Zr_{0.4})_{1-x}M_xNiSn_{0.995}Sb_{0.005}$ (M= V, Nb and Ta) with x= 0%, 0.2%, 0.5% and 1.0%.

| Composition | $S_{RT}$ (μV/K) | n ($10^{19}$cm$^{-3}$) | $\mu_H$ (cm$^2$/(V*s)) | $m^*$ ($m_0$) |
|---|---|---|---|---|
| $Hf_{0.6}Zr_{0.4}NiSn_{0.995}Sb_{0.005}$ | -189 | 1.2 | 38 | 2.1 |
| $(Hf_{0.6}Zr_{0.4})_{0.998}V_{0.002}NiSn_{0.995}Sb_{0.005}$ | -191 | 1.2 | 37 | 2.2 |
| $(Hf_{0.6}Zr_{0.4})_{0.995}V_{0.005}NiSn_{0.995}Sb_{0.005}$ | -198 | 1.1 | 36 | 2.6 |
| $(Hf_{0.6}Zr_{0.4})_{0.99}V_{0.01}NiSn_{0.995}Sb_{0.005}$ | -205 | 1.1 | 34 | 2.7 |
| $(Hf_{0.6}Zr_{0.4})_{0.998}Nb_{0.002}NiSn_{0.995}Sb_{0.005}$ | -184 | 1.3 | 36 | 2.1 |
| $(Hf_{0.6}Zr_{0.4})_{0.995}Nb_{0.005}NiSn_{0.995}Sb_{0.005}$ | -181 | 1.3 | 36 | 2.2 |
| $(Hf_{0.6}Zr_{0.4})_{0.99}Nb_{0.01}NiSn_{0.995}Sb_{0.005}$ | -174 | 1.4 | 35 | 2.1 |
| $(Hf_{0.6}Zr_{0.4})_{0.998}Ta_{0.002}NiSn_{0.995}Sb_{0.005}$ | -183 | 1.3 | 36 | 2.2 |
| $(Hf_{0.6}Zr_{0.4})_{0.995}Ta_{0.005}NiSn_{0.995}Sb_{0.005}$ | -177 | 1.4 | 35 | 2.1 |
| $(Hf_{0.6}Zr_{0.4})_{0.99}Ta_{0.01}NiSn_{0.995}Sb_{0.005}$ | -171 | 1.5 | 36 | 2.1 |

Thermal conductivity measurements were performed on the n-type $(Hf_{0.6}Zr_{0.4})_{0.99}M_{0.01}NiSn_{0.995}Sb_{0.005}$ (M = V, Nb and Ta) and $Hf_{0.6}Zr_{0.4}NiSn_{0.995}Sb_{0.005}$ samples. Among each group of dopant, the 1% doped sample was chosen because it showed the highest PF. In addition, the higher amount of dopants introduces more mass fluctuation in the system, and thus helps to reduce the lattice thermal conductivity.

Thermal conductivity results are shown in Figure 3(a). It can be seen that the value of the thermal conductivity decreases in each of the doped sample. The lattice thermal conductivity results calculated from the relation $\kappa_L = \kappa - \kappa_e$ are shown in Figure 3(b). It shows that the lattice thermal conductivity decreases after replacing (Hf, Zr) with (V, Nb, Ta).

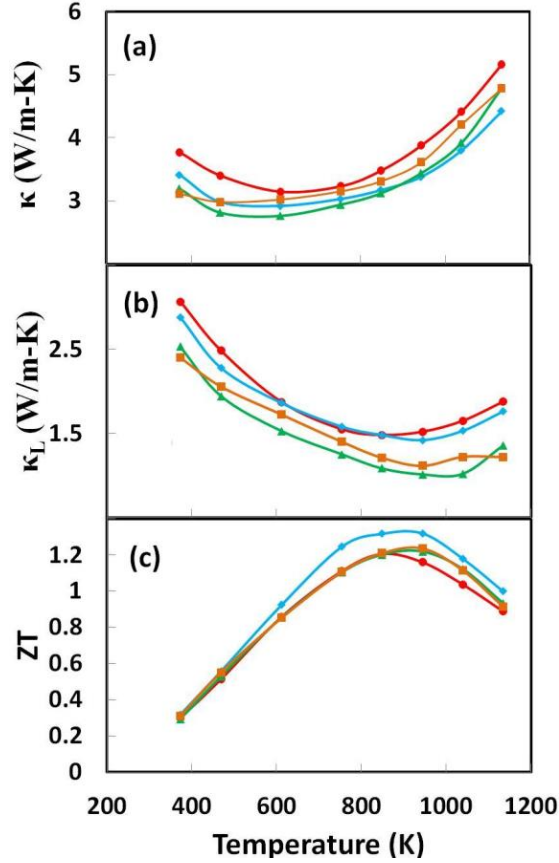

FIG. 3. (a) Thermal conductivity ($\kappa$), (b) lattice thermal conductivity ($\kappa_L$), and (c) ZT of n-type $Hf_{0.6}Zr_{0.4}$ $NiSn_{0.995}Sb_{0.005}$ (red circle), and $(Hf_{0.6}Zr_{0.4})_{0.99}M_{0.01}NiSn_{0.995}Sb_{0.005}$ (M = V, Nb and Ta), where M = V (blue rhombus), M = Nb (green triangle), M = Ta (orange square).

Despite similar power factors in 1% V-doped sample and undoped sample, the figure of merit *ZT* increases due to the decrease of thermal conductivity, as shown in Figure 3(c). The maximum *ZT* value for $(Hf_{0.6}Zr_{0.4})_{0.99}V_{0.01}NiSn_{0.995}Sb_{0.005}$ is about 1.3 at 850 K. In contrast, the maximum *ZT* values of $(Hf_{0.6}Zr_{0.4})_{0.99}Nb_{0.01}NiSn_{0.995}Sb_{0.005}$ and $(Hf_{0.6}Zr_{0.4})_{0.99}Ta_{0.01}NiSn_{0.995}Sb_{0.005}$ are about 1.2.

In summary, a systematic investigation of effect of vanadium, niobium, and tantalum dopants in n-type $Hf_{0.6}Zr_{0.4}NiSn_{0.995}Sb_{0.005}$ half-Heusler alloy was reported. The presence of V resonant states was found to increase both electrical resistivity and Seebeck coefficient. In contrast, Nb and Ta acted as normal dopants, as evident by the decrease in both resistivity and Seebeck coefficient. A high *ZT* of 1.3 was observed near 850K for *n*-type $(Hf_{0.6}Zr_{0.4})_{0.99}V_{0.01}NiSn_{0.995}Sb_{0.005}$ alloys. The increase in *ZT* of V-doped alloys concomitant with the reduction in carrier density and increase in effective band mass was consistent with a bandstructure mechanism of *ZT* enhancement associated with the emergence of resonant states.